\documentclass{ws-procs961x669}            
\begin{document}
\title{Recent results from the Pierre Auger Observatory}

\author{E. Roulet$^*$ for the Pierre Auger Collaboration}

\address{Centro At\'omico Bariloche\\ Av. Bustillo 9500, Bariloche, 8400, Argentina\\
  $^*$E-mail: roulet@cab.cnea.gov.ar\\
Full author list: \url{ https://www.auger.org/archive/authors_2021_07.html}}

\begin{abstract}
The Pierre Auger Observatory has by now achieved an exposure of order $10^5~{\rm km}^2$~sr~yr, exploring about 85\% of the sky. In this talk, I will review some of the recent results, including the detailed measurements of the features in the cosmic ray spectrum, the study of the anisotropies in the cosmic ray arrival directions both at large and intermediate angular scales, the inferred mass composition, and multimessenger searches.
\end{abstract}

\keywords{cosmic rays}

\bodymatter
\section*{}

\begin{figure}[b]
    \centering
    \includegraphics[width=0.55\textwidth]{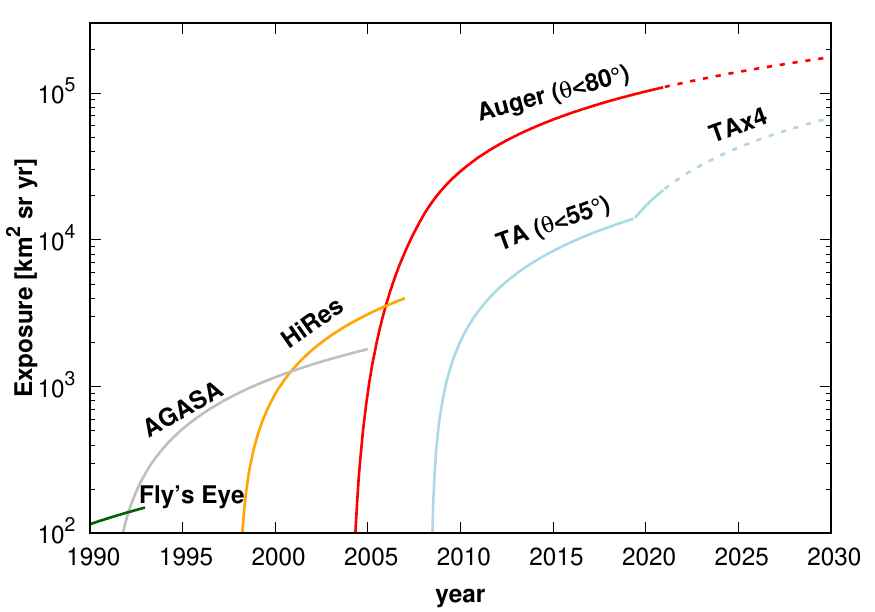}
   \caption{Exposure vs. time of ultrahigh-energy air shower arrays. The previous experiments (Fly's Eye, HiRes and AGASA) as well as the Auger and Telescope Array exposures used in anisotropy searches at the highest energies are shown, together with the extrapolations to the next years involving the enlarged TAx4 upgrade. }
    \label{fexposure}
\end{figure}

\begin{figure}[t]
    \centering
    \includegraphics[width=0.6\textwidth]{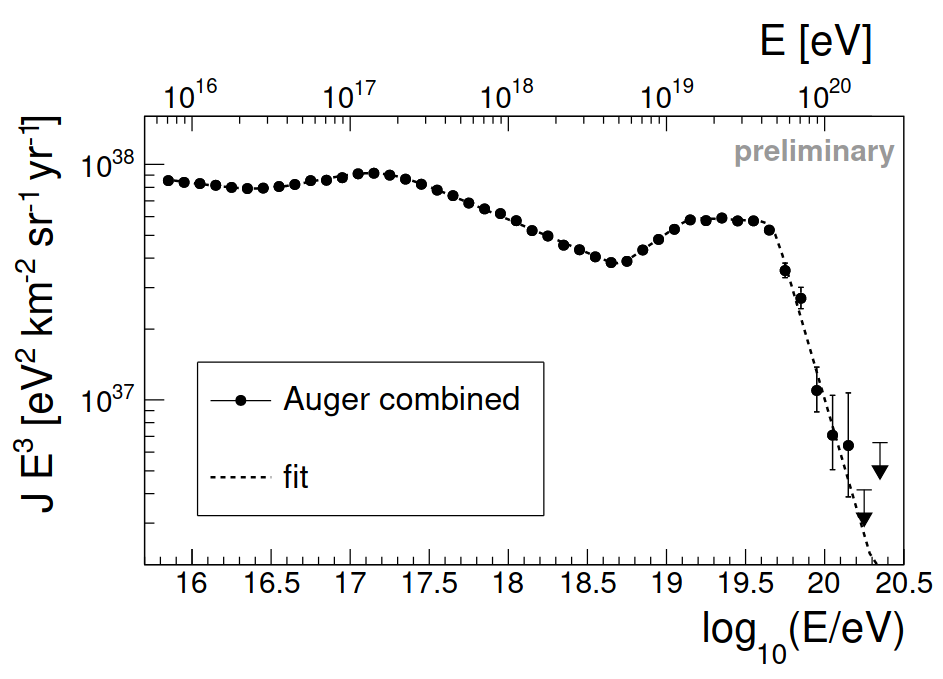}
   \caption{Combined spectrum from measurements obtained with different techniques together with a broken power-law fit (see ref.\citenum{spicrc21} for details). }
    \label{fspectrum}
\end{figure}

The Pierre Auger Observatory has been operational since January 2004 and has by now reached an accumulated exposure of order $10^5$~km$^2$\,yr\,sr (see Fig.~\ref{fexposure}). This enormous increase in the number of detected cosmic rays (CRs) with ultrahigh energies, by more than an order of magnitude with respect to previous observations, has allowed measuring the CR spectrum with unprecedented detail \cite{spicrc21}, firmly establishing the suppression at the highest energies, above 47~EeV, as well as the various features of the spectrum, including the discovery of a softening at 14~EeV, the observation of the ankle at 5~EeV, the second-knee at 0.16~EeV, and the low-energy ankle at 28~PeV (see Fig.~\ref{fspectrum}).

\begin{figure}[t]
    \centering
    \includegraphics[width=0.49\textwidth]{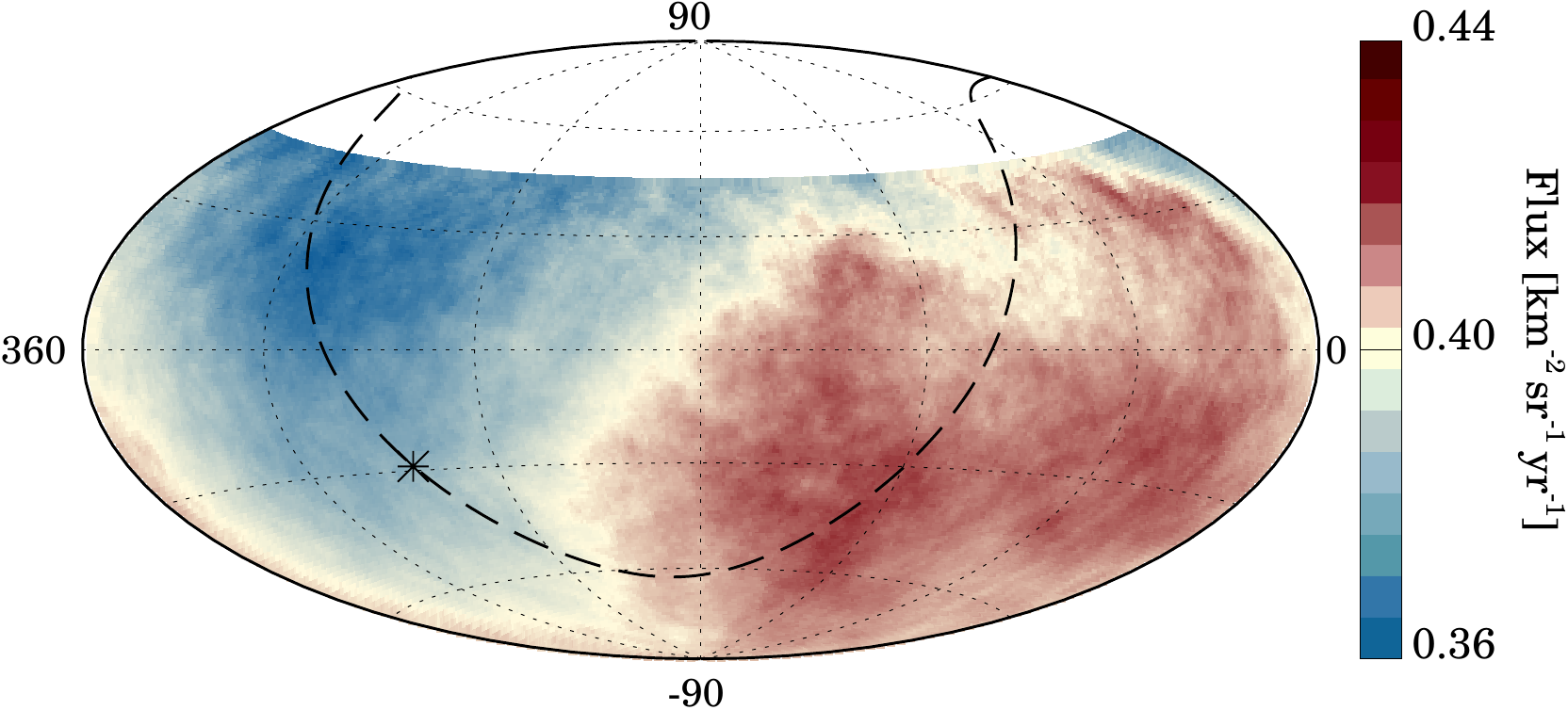}\includegraphics[width=0.49\textwidth]{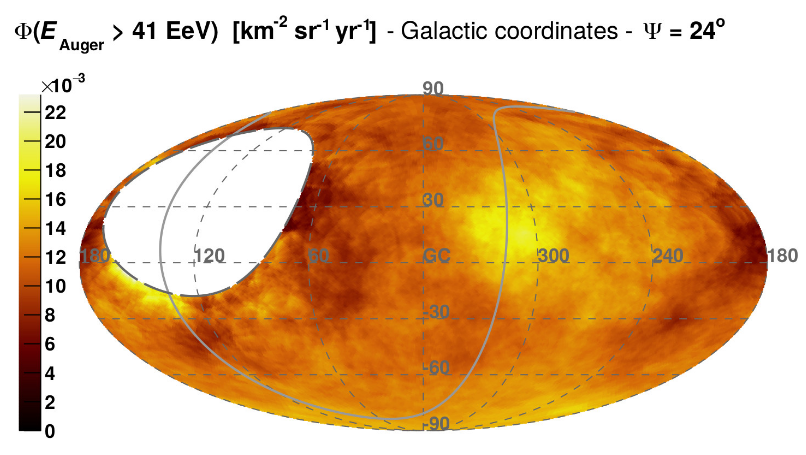}
   \caption{CR flux map above 8~EeV averaged on windows of 45$^\circ$ radius (left panel, in Equatorial coordinates) and above 41~EeV averaged of 24$^\circ$ radius windows (right panel, Galactic coordinates). }
    \label{fmaps}
\end{figure}

By observing  events with zenith angles up to $80^\circ$, i.e., covering 85\% of the sky, it also became possible to determine  with high significance an anisotropy in the CR arrival direction distribution at energies above 8~EeV, which has a characteristic  dipolar pattern with an amplitude of about 7\% \cite{dipole}. The direction of this anisotropy, with the reconstructed dipole pointing about 115$^\circ$ away from the direction towards the Galactic center, provides a clear indication supporting the extragalactic origin of the CRs at these energies (see left panel of Fig.~\ref{fmaps}). Considering several energy bins above 8~EeV, the dipole amplitude is found to actually increase with energy, and the direction of the fitted dipole in all bins lies not far from the direction of the outer spiral arm of the Galaxy \cite{dipolebins}. At energies below 8~EeV, no significant anisotropies were observed, so that relevant upper bounds on the equatorial dipole amplitudes were set down to energies of  0.03~EeV, and interestingly the right ascension phase of the flux modulations below about 1~EeV is not far from the right ascension of the Galactic center \cite{LSaniso}.  Hints of anisotropies at intermediate angular scales, of order 20$^\circ$, have been obtained at energies above about 40~EeV, with the main excess lying around the direction towards the nearby radiogalaxy Centaurus~A (right panel of Fig.~\ref{fmaps}). This radiogalaxy then provides a possible candidate source of ultrahigh-energy CRs, although other potential CR sources are also located in the same region of the sky, such as the starburst galaxies NGC4945 and M83.
A likelihood ratio test comparing the CR arrival directions above different threshold energies with a catalog of nearby starburst galaxies, flux-weighted and smeared, reveals a maximum significance above a threshold of 38~EeV for an angular scale of a top-hat window of 25$^\circ$ \cite{anisoa32}.

\begin{figure}[t]
    \centering
    \includegraphics[width=0.85\textwidth]{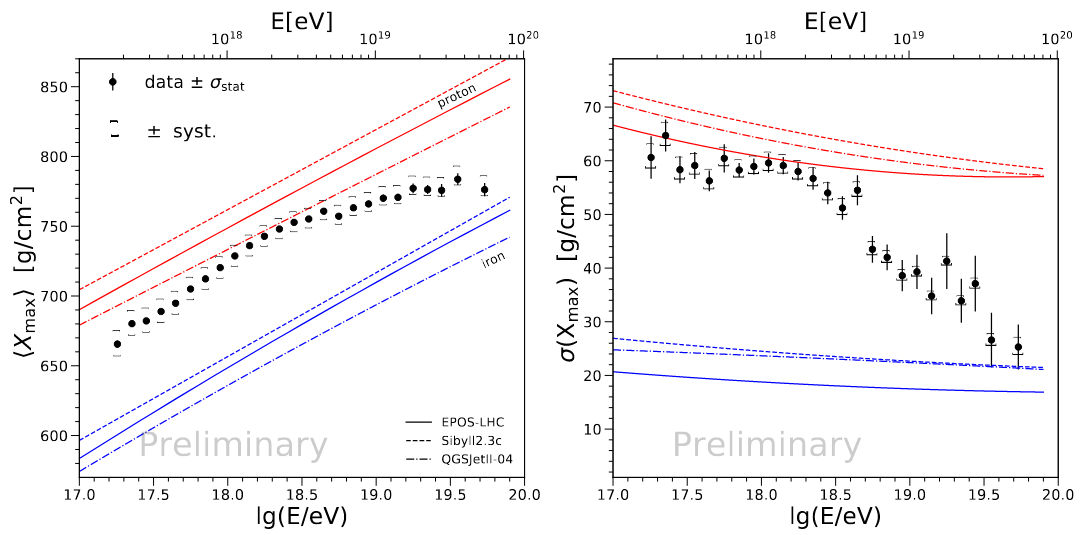}
   \caption{Mean $X_{\rm max}$ and its dispersion as a function of energy, as measured with the Auger fluorescence detectors. }
    \label{fxmax}
\end{figure}

The hybrid nature of the Auger Observatory, with both surface detectors (SD) and fluorescence detectors (FD), allows it to perform sensitive measurements of the mass composition of the primary CRs. The SD 
consists of instrumented water tanks that measure the Cherenkov light emitted by relativistic secondary charged particles crossing them and is hence sensitive to both the muonic and the electromagnetic components of the showers reaching ground level. The FD telescopes measure instead the longitudinal development in the atmosphere of the CR induced showers by observing, with a $\sim 15$\% duty cycle, the light emitted by the N$_2$ air molecules that get excited by the passage of the electromagnetic component of the shower. The FD allows for an almost calorimetric measurement of the primary energy, which is also used to calibrate the SD energy assignment. Moreover, the atmospheric depth at which the showers reach their maximum development, $X_{\rm max}$, provides crucial information about the mass composition of the CRs, because lighter primaries give rise to more penetrating showers than heavier ones since these last can be considered as a superposition of many lower energy showers induced by the individual nucleons.
The measurements of the distribution of $X_{\rm max}$ values allowed us to establish \cite{compositionFD} that at energies of a few EeV the average CR masses are light, being dominated by H and He primaries, and they become increasingly heavier as the energy is further increased, lying in the ballpark of the expectations from primary N nuclei at energies of 30~EeV and possibly even heavier at higher energies (the actual masses inferred depend, however, on the hadronic interaction model adopted in the analysis). The narrow spread in the fluctuations on the values of $X_{\rm max}$  measured above the ankle energy also implies that there is little admixture between elements with disparate masses at these energies. These measurements are depicted in Fig.~\ref{fxmax}. 

A combined fit to the spectrum and anisotropy measurements within an astrophysical scenario with equal luminosity sources having power-law spectra with rigidity dependent cutoffs has been performed in ref.~\citenum{combinedfit}.

Besides the measurement of $X_{\rm max}$ with FD, also the SD has been used to obtain composition information \cite{compositionSD}, exploiting the rise-times of the measured signals in the triggered detectors. Given that the shower muons travel straight from their production points high in the atmosphere, while the electromagnetic component of photons, electrons, and positrons have a more diffusive propagation in the atmosphere, the muonic signal is expected to be more concentrated in the early part of the shower front while the electromagnetic one will have a larger spread. Hence, shorter rise times are expected in the signals from showers with a more abundant muonic component, as is the case for the showers induced by heavier nuclei. These measurements can then be correlated with the FD measurements of the depth of shower maxima. Given the larger statistics obtained with SD, which has a 100\% duty cycle, it has become possible to obtain more detailed information about the composition at the highest energies, actually splitting the highest energy FD bin in four SD bins. In this way the trend towards a heavier composition for increasing masses is confirmed with SD up to the highest energies.

\begin{figure}[t]
    \centering
    \includegraphics[width=0.49\textwidth]{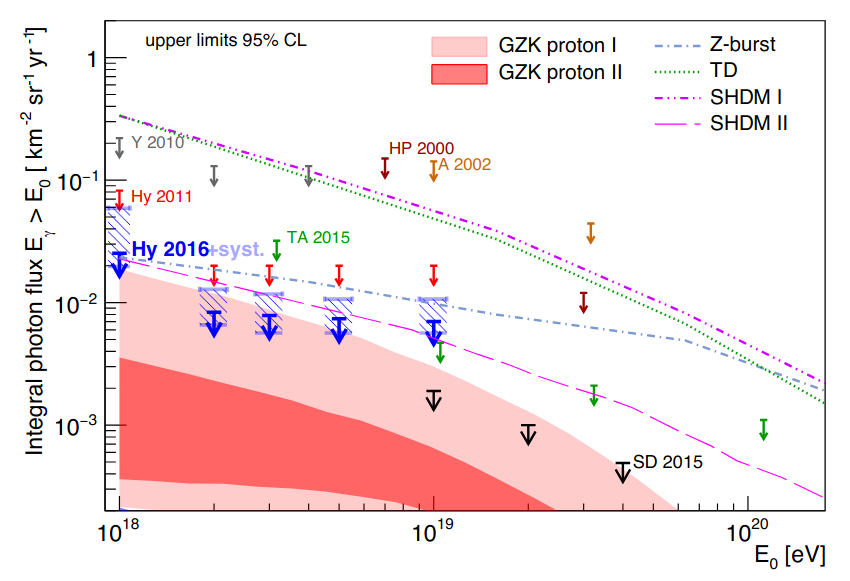}  \includegraphics[width=0.49\textwidth]{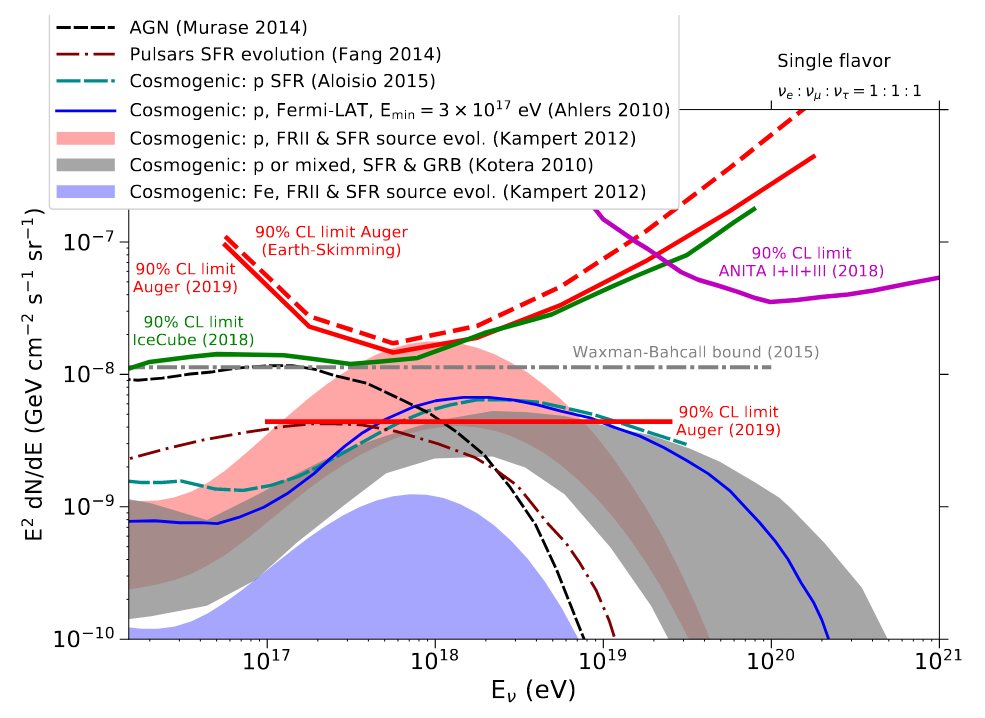}
   \caption{Bounds on photon fluxes (left) and neutrino fluxes (right), compared to some model expectations. }
    \label{fgammanu}
\end{figure}

Another search performed has been to look for a component of photon-induced showers, which would be very poor in muons, and this allowed to set stringent bounds on the primary photon fluxes above EeV energies \cite{photons}. The upper bounds strongly disfavor exotic scenarios for the production of UHECRs, such as those involving superheavy particle decays (see left panel in Fig.~\ref{fgammanu}).

Also ultrahigh-energy neutrinos have been searched through the study of inclined showers \cite{nubounds}. Very inclined showers produced by ordinary nuclear-induced cascades, which initiate high in the atmosphere, would have their electromagnetic component totally attenuated by the time they reach ground level. On the other hand,  neutrino-induced showers can be initiated at any depth in the atmosphere, and those originating close to the detector will have a significant electromagnetic component, looking then `younger' than the hadronic ones, what could be exploited to single them out from the background. Actually, the most sensitive neutrino search is that looking for showers produced by Earth-skimming tau neutrinos interacting in the crust of the Earth and producing tau leptons that manage to exit from the soil before decaying to produce the atmospheric shower that may be observed. The most sensitive energy range for these searches is  around the EeV, which just happens to be the range where the fluxes of cosmogenic neutrinos are expected to peak. In particular, cosmogenic neutrinos could arise from the photo-pion production taking place when CR protons with energies exceeding the GZK threshold of about 40~EeV interact with CMB photons, and the neutrinos produced in the charged pion decays have typical energies of about $E_p/20$. No neutrino candidates have yet been observed in these searches, allowing us to set bounds that start to constrain some of the most optimistic scenarios for the production of cosmogenic neutrinos, which are  those with protons dominating the CR fluxes at the highest energies. However, more realistic scenarios with a predominantly heavier composition at the highest energies predict fluxes below the present bounds (see right panel of Fig.~\ref{fgammanu}).

Searches of neutrinos in association with the first neutron star binary merger detected by the LIGO-Virgo Collaboration also led to negative results \cite{nsmerger}, in spite of the lucky coincidence that the event just happened to be a few degrees below the horizon, a situation in which the sensitivity to Earth-skimming tau neutrinos is largely enhanced. It is interesting to note that the model predictions for the case in which the jet of this closeby merger had been aligned with the line of sight were not very far from the Auger sensitivity, but the lack of an observed signal is not unexpected, given the slight misalignment of the jet in this particular merger. Also searches in the direction towards the blazar TXS 0506+056, from which a $\sim 300$~TeV neutrino was observed by IceCube, revealed no neutrino candidate, allowing to constrain the neutrino flux from this object at the highest energies \cite{txs}.

The Pierre Auger Observatory is at present undergoing an upgrade, consisting in adding 4~m$^2$ scintillators on top of each SD and using faster electronics,  which in combination with the water-Cherenkov detectors should allow for a cleaner separation between the electromagnetic and muonic components of the showers. This separation should make it possible to obtain composition information on an event by event basis with the SD detector, allowing, for instance, to improve the sensitivity of the anisotropy studies by restricting them to just the lighter CR component (which is less deflected by the Galactic and extragalactic magnetic fields). Also radio antennas are being added to each SD detector to observe the radio emission from the electromagnetic component of the showers, which should help in particular to study the composition of the inclined showers.  This new phase of the Observatory should allow for improved studies of all the topics just discussed, as well as to better handle the still remaining open issues.

\end{document}